\documentclass[aps,10pt,nofootinbib,preprint]{revtex4-1}
\usepackage[T1]{fontenc}
\usepackage{hyperref}
\usepackage{xcolor}
\usepackage{graphicx}
\usepackage{amsmath,amssymb}
\usepackage{bm}

\begin{document}
	\title{\boldmath Stationary bubbles: information loss paradox?}
	
	\author{{Guillem Dom\`enech$^*$} and {Misao Sasaki$^\dag$}}
	\address{Yukawa Institute for Theoretical Physics,\\ 
		Kyoto University, Kyoto 606-8502, Japan\\
		$^*$guillem.domenech{}@{}yukawa.kyoto-u.ac.jp\\
		$\dag$misao{}@{}yukawa.kyoto-u.ac.jp}
	\date{\today}
	\preprint{YITP-16-12}

	\begin{abstract}
    The main purpose of this work is to build classically stationary bubbles, within the thin-shell formalism, which are unstable under quantum effects; they either collapse into a black hole or expand. Thus, the final state can be thought of a superposition of geometries. We point out that, from a quantum mechanical point of view, there is no issue with a loss of information in such configuration. A classical observer sees a definite geometry and, hence, finds an effective loss of information. Although it does not cover all possible cases, we emphasise the role of semi-classical gravitational effects, mediated by instatons, in alleviating/solving the information loss paradox.\\\\\\\\\\\\\\\\\\\\\\
    Note: Prepared for the Proceedings for the 2nd LeCosPA Symposium:
    Everything about Gravity.
	\end{abstract}
	
	\maketitle

	\section{Introduction \label{intro}}
	Black holes are truly interesting objects in General Relativity but not free from problems. In particular, the information loss paradox\cite{Hawking:1976ra} is a long standing question without a definitive answer. It has been shown that one cannot hold on to
	general relativity, semi-classical quantum field theory, area-entropy relation, 
	and unitarity all at the same time \cite{Yeom:2008qw,Almheiri:2012rt}.
	
%
%
%
%
%
%
%

	Although there are several proposals with their own share of issues\cite{Hawking:2005kf,Susskind:1993if,Yeom:2008qw,Almheiri:2012rt,Unruh:1995gn,Hwang:2012nn,Ashtekar:2005cj,Page:2013mqa,Chen:2014jwq,Sasaki:2014spa,Hawking:2016msc}, here we tackle the information loss paradox from the euclidean path integral approach and the wave function of the universe\cite{Hartle:1983ai}. There, one sums over all possible geometries and configurations, formally given by
	\begin{align}
	\psi(\phi_i,g_i|\phi_f,g_f)=\int {\cal D}\phi{\cal D}g	~{\rm e}^{iS[\phi,g]}\,,	
	\end{align}
	but in practise one cannot deal with the whole path integral. Nevertheless, there is an interesting proposal by Maldacena \cite{Maldacena:2001kr} and Hawking \cite{Hawking:2005kf} who argued that if one of those histories has a trivial topology, i.e. no horizon nor singularity, then information is conserved in the whole wave function. However, a classical observer who sees a definite geometry experiences an \textit{effective loss of information}. In this work, we construct concrete examples supporting this point of view. Although one cannot build them in general, we show that there is a wide range of parameters that allow such configurations.
	
	We have in mind the picture of a star collapse, where a stable star suffers for any reason a phase transition and undergoes gravitational collapse, perhaps forming a black hole. We model this behaviour within the thin shell formalism and look for a stationary bubble solutions. To do that a simple tension does not suffice and thus we modify it at high energies regime to obtain the desired solution.
	
\begin{figure}[t]
		\centering
		\includegraphics[width=0.7\columnwidth]{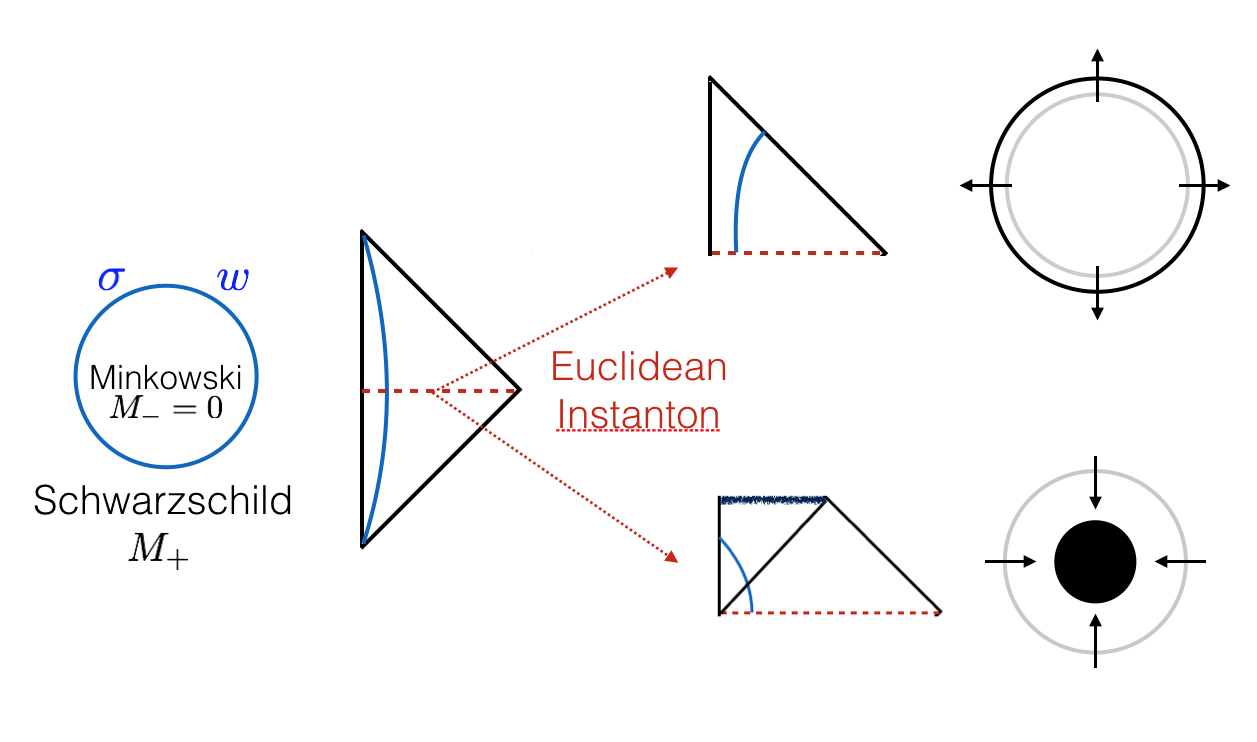}
		\caption{Rough sketch of the set up. A stationary thin shell, constant $r$ blue curve in the left hand side, tunnels into a black hole, right-down, or expands to infinity, right-up. Those two outcomes are possible and hence the final state is a combination of two geometries.}
		\label{fig:1}
	\end{figure}

	The idea of this work is as follows. As schematically shown in figure \ref{fig:1}, we want to build a stationary bubble, as much as a toy model for a star collapse with a regular interior, which the final geometry, black hole or expanding shell, is due to semi-classical effects. Thus, in section \ref{shell} we review the thin-shell formalism, in section \ref{stationary} we build a stationary thin shell solution and discuss the implications to the information loss paradox. Finally, in section \ref{conc} we summarize and conclude our work.
		
	\section{Thin-shell junction conditions\label{shell}}
	
	By virtue of Israel junction conditions \cite{Israel:1966rt} one can glue two solutions of Einstein equations, allowing for a thin shell which accounts for any jump of the extrinsic curvature at the junction. In particular, assume a spherically symmetric system where the metric is given by
	\begin{eqnarray}
		\label{eq:metric}
		ds_{\pm}^{2}= - f_{\pm}(R) dT^{2} + \frac{1}{f_{\pm}(R)} dR^{2} + R^{2} d\Omega^{2}\qquad \left(f_{\pm}(R) = 1 - \frac{2M_{\pm}}{R} + \frac{R^{2}}{\ell_{\pm}^{2}}\right)\,,
	\end{eqnarray}
	where $\pm$ refer to outside ($R>r$) and inside ($R<r$) the shell, $M_\pm$ is the mass parameter and $\ell_\pm$ is the AdS radius. Then
    the junction condition yields
    \begin{eqnarray}\label{eq:junc}
    	\epsilon_{-} \sqrt{\dot{r}^{2}+f_{-}(r)} - \epsilon_{+} \sqrt{\dot{r}^{2}+f_{+}(r)} = 4\pi r \sigma\,,
    \end{eqnarray}
    where $\sigma$ is the tension of a perfect fluid thin shell and $\epsilon_\pm$ is the sign of the extrinsic curvature out/inside respectively. On top of that we have the conservation of energy, relating surface energy density $\sigma$ and pressure $\lambda$,
    which given a constant equation of state $w=\lambda/\sigma$ leads us to
    \begin{eqnarray}
        	\sigma(r)= \frac{\sigma_0}{r^{2(1+w)}}\,.
    \end{eqnarray}
    With this simple set of equations one can find very interesting examples\cite{Adler:2005vn,Mann:2006yu}. In most cases though, a simple choice of the tension yields only two possible outcomes; collapse or expansion. In the next section we show some sufficient requirements for the tension so as to obtain a stationary thin shell or bubble. Before that, let us remind the reader that the situation gets more interesting when semi-classical effects come into play\cite{Farhi:1989yr,Ansoldi:1997hz}.

    \subsection{Effective potential and thin shell tunnelling}
    
       	    \begin{figure}[t]
       	    	\centering
       	    	\includegraphics[width=0.4\columnwidth]{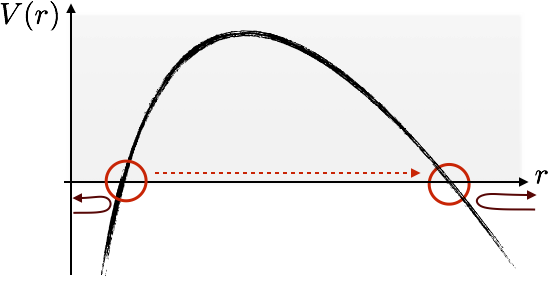}
       	    	\hspace{1cm}
       	    	\includegraphics[width=0.35\columnwidth]{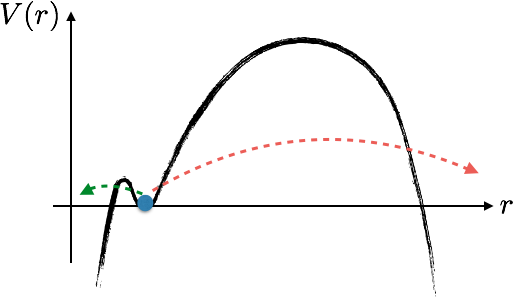}
       	    	\caption{Schematic form of the effective potential. Left hand side shows the general behaviour for true vacuum bubbles in AdS background with a constant thin shell tension. Right hand side shows the desired behaviour we want to achieve: a stationary bubble.}
       	    	\label{fig:2}
       	    \end{figure}
       	    
    One can grasp a more physically intuitive picture by rewriting the junction condition eq.\eqref{eq:junc} as a particle with zero energy moving in a 1-D potential, that is\cite{Blau:1986cw}:
    \begin{eqnarray}
    	\label{eq:form}
    	\dot{r}^{2} + V(r) = 0,\quad
    	V(r) = f_{+}(r)- \frac{\left(f_{-}(r)-f_{+}(r)-16\pi^{2} \sigma^{2} r^{2}\right)^{2}}{64 \pi^{2} \sigma^{2} r^{2}}.
    	\label{eq:form2}
    \end{eqnarray}
    This form is particularly useful to understand the dynamics of the thin shell with radius $r$. For example, for true vacuum bubbles in an AdS background with a constant tension,\cite{Sasaki:2014spa} the potential generally takes the form as in the left side of figure \ref{fig:2}. It is easy to see that the shell either unavoidably collapse or expands. Now, if our model parameters have a region with $V(r)>0$, we have two classically disconnected regions. On one hand, the shell starts from $r=0$, expands, bounces and collapses back to $r=0$. On the other hand, the shell starts from $r=\infty$, collapses, bounces and expands. However, as it is well know from quantum mechanics, a particle can cross a classically forbidden region by a quantum tunnelling. In our case, the thin shell can tunnel,\cite{Ansoldi:1997hz} for example, from an expanding solution from $r=0$ to an expanding solution to $r=\infty$, with a exponentially suppressed yet non-vanishing probability.
%
   	The main drawback of the latter example though, is that the shell starts from a white hole at $r=0$ and thus one could question its initial conditions. For this reason, let us modify the latter example and look for an stationary bubble as an initial condition, see right of figure \ref{fig:2}.
    
    Before ending this section, let us recall that the sign of the extrinsic curvature dictates the location of the thin shell in the Penrose diagram. We require the extrinsic curvature to be positive for $r$ larger than the outer horizon so as to be located at the right hand side of the diagram. This is ensured by requiring $\sqrt{f_--f_+}>4\pi\sigma r$, as one can see from recovering the signs of the extrinsic curvature, that is
    \begin{eqnarray}
    	\beta_{\pm}(r) \equiv \frac{f_{-}(r)-f_{+}(r)\mp 16\pi^{2} \sigma^{2} r^{2}}{8 \pi \sigma r} = \epsilon_{\pm} \sqrt{\dot{r}^{2}+f_{\pm}(r)}\,.
    	\label{eq:extrinsic}
    \end{eqnarray}
    We show this requirement as a blue line in all forthcoming plots.
    
	\section{Stationary bubbles and superposition of geometries\label{stationary}}
	
    \begin{figure}[t]
		   	\centering
		   	\includegraphics[width=0.85\columnwidth]{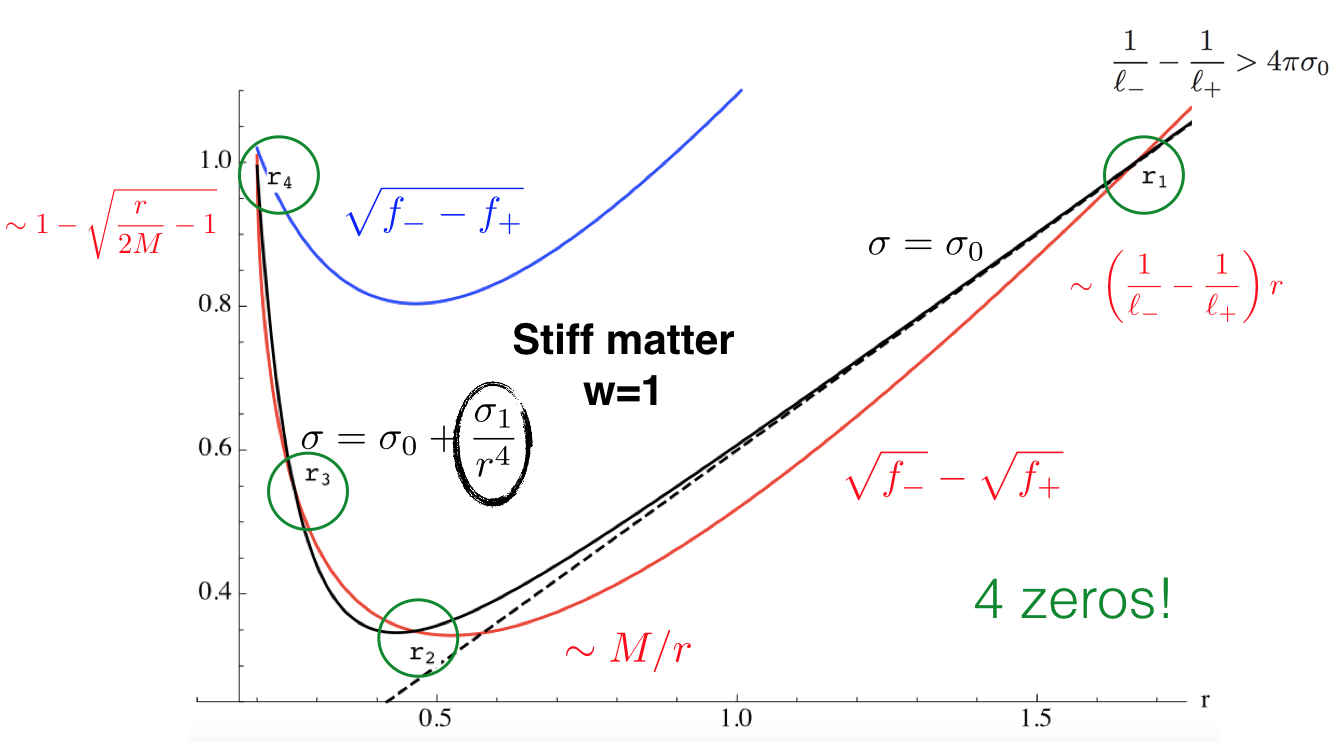}
		   	\caption{Counting zeros of the potential. The red line is $\sqrt{f_-}-\sqrt{f_+}$ and the approximate behaviour in each zone is written in red. Black lines are $4\pi\sigma r$, black dotted for a constant tension and black curve for the modified behaviour $\sim1/r^4$. Any crossing with the blue line indicates a change in the extrinsic curvature sign. The parameters of this plot are as follows ($c=\hbar=G=1$): $M_+=0.1$, $M_-=0$, $\ell_+=\infty$, $\ell_-=2$, $4\pi\sigma_0=0.3$ and $4\pi\sigma_1=0.0067$.}
		   	\label{fig:3}
    \end{figure}
    
	Let us intuitively explain how the tension $\sigma$ should behave at high energies (small $r$) regime so as to achieve a classically stable configuration. For this purpose, it is useful to count the number of zeros in the effective potential, by ploting the junction condition eq.\eqref{eq:junc} setting $\dot r=0$, rather than the effective potential form. For example, for an AdS true vacuum bubble one generally obtains the behaviour shown in figure \ref{fig:3}. Notice that for a constant tension, which corresponds to the black dotted line, we obtain two intersections with the red line and, therefore, the potential has two zeros and takes the form of left figure \ref{fig:2}.
	
	Now, we want to modify this solution so as to obtain four zeros in the effective potential. In other words, we want four intersections with the red line in total. From figure \ref{fig:3}, it is readily seen that the tension must be steep enough in order to overcome the $1/r$ behaviour of the red line curve. We choose to add stiff matter with an equation of state $w=1$, which leads to a $1/r^4$ behaviour of the tension. See the black line in figure \ref{fig:3}. This is obviously not the only possibility but the simplest modification, as far as we understand.

\begin{figure}[t]
    	\centering
    	\includegraphics[width=0.45\columnwidth]{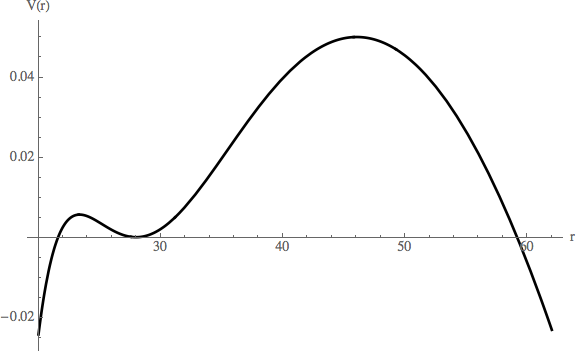}\hspace{0.5cm}
    	\includegraphics[width=0.45\columnwidth]{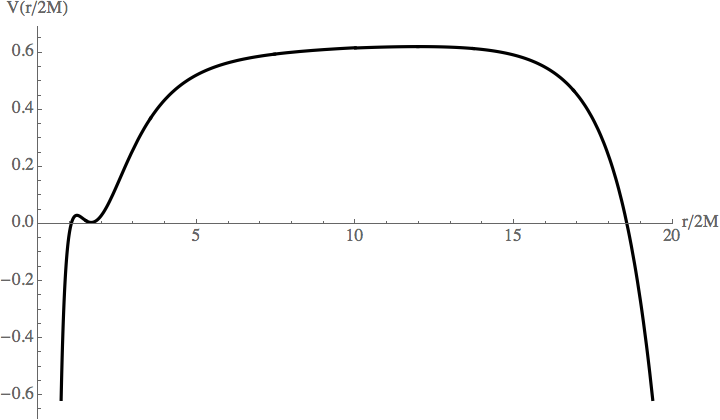}
    	\caption{Concrete examples of the potential. On the left hand side the parameters are: $M_+=10$, $M_-=0$, $\ell_+=\infty$, $\ell_-=40$, $4\pi\sigma_0=0.016$ and $4\pi\sigma_1=5230$. On the right hand side we chose: $\ell_\pm=\infty$, $\bar\sigma_0=0.87$, $\mu_{1}=1/2$, $\mu_{2}=1/3$, $\bar{r}_{1}=1$ and $\bar{r}_{2}=20$.}
    	\label{fig:4}
    \end{figure}

    The same procedure can be applied in the case of a thin shell in a Minkowski background, although one must tune more the tension\cite{Sasaki:2014spa}. In particular, we use:
    \begin{align}
    	w(\bar r)=\left\{
    	\begin{array}{ccc}
    		\frac{1}{2}\left(\tanh\left[\mu_1\left(\bar{r}_1-\bar{r}\right)\right] +1\right)& \quad \bar{r}<\bar{r}_m
    		\\
    		\frac{1}{6}\left(\tanh\left[\mu_2\left(\bar{r}-\bar{r}_2\right)\right] +1\right)& \quad \bar{r}>\bar{r}_m
    	\end{array}
    	\right.,
    \end{align}
    where $\bar{r}=r/2M$, $\bar{\sigma}\equiv8\pi M\sigma$, $\mu_i$ and $r_i$ respectively control the steepness and position of the transition, and $r_m$ is the matching radius in an intermediate region. One can easily find a parameter region where three or four zeros are obtained and, therefore, find a stationary solution. A concrete example is shown in figure \ref{fig:4}.

    So far we showed examples of stationary bubbles, which we think of a toy model of a star collapse. Now, it is time to discuss the implications to the information loss paradox. 
    In fact, one can numerically compute the probabilities\cite{Ansoldi:1997hz} and find that the probability of tunnelling into a black hole is exponentially higher than the probability to tunnel to an expanding shell. Still, the latter is non-vanishing and therefore there is a path in the whole wave function of the universe where the information is recovered. That implies that a classical observer who sees a definite geometry, the most probable one, finds an \textit{effective loss of information}.

	\section{Conclusion \label{conc}}
	
	We studied the information loss paradox in the context of euclidean path integrals. Using the thin shell formalism, we built classically stable stationary bubbles. We showed the existence of such examples for true vacuum bubbles in asymptotically flat and AdS backgrounds and also in Minkowski spacetime. They could be thought of as a toy model of a star collapse. Afterwards, we considered that such bubbles are not stable under quantum effects and eventually they tunnel, mediated by instantons, either to a collapsing shell that forms a black hole or to an expanding shell that expands all the way to infinity. In terms of the wave function, the final geometry is a composition of two possible final states. However, a classical observer only sees a definite geometry and, most likely, finds a black hole configuration since it is exponentially favoured with respect to the expanding solution. Thus, a classical observer might find an effective loss of information, while the whole information is encoded in the wave function of the universe. Although our work is based in a collection of examples, we want to emphasise that in such cases the role of semi-classical effects in solving the information loss paradox is essential. It might be helpful, for example, in cases such as eternal black holes or extremal charged black holes. 
	
	\section*{Acknowledgments}
			GD and MS would like to thank P.Chen and D.h.Yeom for many fruitful discussions. This work was supported in part by MEXT KAKENHI Grant Number 15H05888.


\begin{thebibliography}{200}
	\bibitem{Hawking:1976ra}
	S.\,W.\,Hawking,
	Phys.\ Rev.\  D {\bf 14}, 2460 (1976).
	
	\bibitem{Hawking:2005kf}
	S.\,W.\,Hawking,
	Phys.\ Rev.\  D {\bf 72}, 084013 (2005)
	[arXiv:hep-th/0507171];
	S.\,W.\,Hawking,
	arXiv:1401.5761 [hep-th].

	\bibitem{Susskind:1993if}
	L.\,Susskind, L.\,Thorlacius and J.\,Uglum,
	Phys.\ Rev.\ D {\bf 48} (1993) 3743
	[hep-th/9306069].
	
	\bibitem{Yeom:2008qw}
	D.\,Yeom and H.\,Zoe,
	Phys.\ Rev.\  D {\bf 78}, 104008 (2008)
	[arXiv:0802.1625 [gr-qc]];
	P.\,Chen, Y.\,C.\,Ong and D.\,Yeom,
	JHEP {\bf 1412}, 021 (2014)
	[arXiv:1408.3763 [hep-th]].
	
	\bibitem{Almheiri:2012rt} 
	A.\,Almheiri, D.\,Marolf, J.\,Polchinski and J.\,Sully,
	JHEP {\bf 1302}, 062 (2013)
	[arXiv:1207.3123 [hep-th]];
	A.\,Almheiri, D.\,Marolf, J.\,Polchinski, D.\,Stanford and J.\,Sully,
	JHEP {\bf 1309}, 018 (2013)
	[arXiv:1304.6483 [hep-th]].
	
	\bibitem{Unruh:1995gn}
	W.\,G.\,Unruh and R.\,M.\,Wald,
	Phys.\ Rev.\ D {\bf 52} (1995) 2176
	[hep-th/9503024].
	
	\bibitem{Hwang:2012nn} 
	D.\,Hwang, B.\,-H.\,Lee and D.\,Yeom,
	JCAP {\bf 1301}, 005 (2013)
	[arXiv:1210.6733 [gr-qc]];
	W.\,Kim, B.\,-H.\,Lee and D.\,Yeom,
	JHEP {\bf 1305}, 060 (2013)
	[arXiv:1301.5138 [gr-qc]].
	
	
	\bibitem{Ashtekar:2005cj}
	A.\,Ashtekar and M.\,Bojowald,
	Class.\ Quant.\ Grav.\  {\bf 22} (2005) 3349
	[gr-qc/0504029];
	P.\,M.\,Ho,
	arXiv:1510.07157 [hep-th].
	
	
	
	\bibitem{Page:2013mqa}
	D.\,N.\,Page,
	JCAP {\bf 1406} (2014) 051
	[arXiv:1306.0562 [hep-th]];
	P.\,Chen, Y.\,C.\,Ong, D.\,N.\,Page, M.\,Sasaki and D.\,Yeom, arXiv:1511.05695 [hep-th].
	
		
	\bibitem{Chen:2014jwq}
	P.\,Chen, Y.\,C.\,Ong and D.\,h.\,Yeom,
	Phys.\ Rept.\  {\bf 603} (2015) 1
	[arXiv:1412.8366 [gr-qc]];
	W.\,Y.\,Wen and S.\,Y.\,Wu,
	Eur.\ Phys.\ J.\ C {\bf 75} (2015) 12,  608
	[arXiv:1509.01305 [hep-th]];
	Y.\,C.\,Ong,
	JCAP {\bf 1504} (2015) 04,  003
	[arXiv:1503.01092 [gr-qc]].
	
	\bibitem{Sasaki:2014spa} 
	M.\,Sasaki and D.\,Yeom,
	JHEP {\bf 1412}, 155 (2014)
	[arXiv:1404.1565 [hep-th]];
	P.\,Chen, G.\,Dom{\`e}nech, M.\,Sasaki and D.\,h.\,Yeom,
	arXiv:1512.00565 [hep-th];
	P.\,Chen, Y.\,C.\,Hu and D.\,h.\,Yeom,
	arXiv:1512.03914 [hep-th].
	
	\bibitem{Hawking:2016msc}
	S.\,W.\,Hawking, M.\,J.\,Perry and A.\,Strominger,
	arXiv:1601.00921 [hep-th].
	A.\,Averin, G.\,Dvali, C.\,Gomez and D.\,Lust,
	arXiv:1601.03725 [hep-th].
	
	\bibitem{Hartle:1983ai}
	J.\,B.\,Hartle and S.\,W.\,Hawking,
	Phys.\ Rev.\  D {\bf 28}, 2960 (1983).
	
	\bibitem{Maldacena:2001kr}
	J.\,M.\,Maldacena,
	JHEP {\bf 0304}, 021 (2003)
	[arXiv:hep-th/0106112].
	
%

	
		
	
	\bibitem{Israel:1966rt}
	W.\,Israel,
	Nuovo Cim.\ B {\bf 44S10} (1966) 1
	
	
	\bibitem{Adler:2005vn}
	R.\,J.\,Adler, J.\,D.\,Bjorken, P.\,Chen and J.\,S.\,Liu,
	Am.\ J.\ Phys.\  {\bf 73}, 1148 (2005)
	[gr-qc/0502040].
	\bibitem{Mann:2006yu}
	R.\,B.\,Mann and J.\,J.\,Oh,
	Phys.\ Rev.\ D {\bf 74} (2006) 124016
	[gr-qc/0609094].
	
	\bibitem{Farhi:1989yr}
	E.\,Farhi, A.\,H.\,Guth and J.\,Guven,
	Nucl.\ Phys.\  B {\bf 339}, 417 (1990); 
	W.\,Fischler, D.\,Morgan and J.\,Polchinski,
	Phys.\ Rev.\  D {\bf 42}, 4042 (1990).
		
	\bibitem{Ansoldi:1997hz}
	S.\,Ansoldi, A.\,Aurilia, R.\,Balbinot and E.\,Spallucci,
	Class.\ Quant.\ Grav.\  {\bf 14} (1997) 2727
	[gr-qc/9706081].
	
	\bibitem{Blau:1986cw} 
	S.\,K.\,Blau, E.\,I.\,Guendelman and A.\,H.\,Guth,
	Phys.\ Rev.\ D {\bf 35}, 1747 (1987).
	


	
%
	
\end{thebibliography}
\end{document}